%Paper: hep-th/9211120
%From: michael terhoeven <UNP044@IBM.rhrz.uni-bonn.de>
%Date: Wed, 25 Nov 92 10:09:20 MEZ
%Date (revised): Tue, 01 Dec 92 10:14:10 MEZ

%%%%%%%%%%%%%%%%%%%%%%%%%%%%%%%%%%%%%%%%%%%%%%%%%%%%%%%%%%%%%%%%%%%%%
\magnification=\magstep1  % for the printer 502 use 1000=\magstep0
\vsize=8.4truein
\voffset=0.5truein
\hoffset=0.35truein   % war 0.45
\hsize=6.3truein     % setting the size of a page
\nonfrenchspacing    % different spacings after , ! ? etc
\interlinepenalty=10  % penalty for page-break between lines of one paragraph
\baselineskip=16pt

\font\grosss=cmr7 scaled \magstep4

\font\gros=cmssbx10 scaled \magstep2

\outer\def\unter#1 #2\par{\vskip 0pt plus .1\vsize \penalty-200
     \vskip 0pt plus -.1\vsize \bigskip \vskip \parskip
     \message{#1 #2
       } \vfill\leftline{\gros #1 \bf#2}\nobreak\medskip\noindent}

\def \rrr{{\rm I \! R}}
\def \ccc{{\rm C \kern-5.5pt I \ }}
\def \zzz{{\rm Z \!\! Z}}

\def \ccp{{\rm C \kern-5.5pt I\, P}}

\def \mod{{\rm\ mod\ }}

\def \eins{{\rm 1 \kern-2.8pt I }}

          % { {\cal L} \kern-10pt - }

\def \>>{\rangle}      %{\mkern-5mu >}
\def \tchi{\chi \kern-5.6pt\chi }

\def\frac#1/#2{\leavevmode\kern.1em
              \raise.5ex\hbox{$\the\scriptfont0 #1$}\kern-.1em
              /\kern-.05em\lower.4ex\hbox{$\the\scriptfont0 #2$}}

\def \sqr#1#2{{\vcenter{\vbox{\hrule height.#2pt
            \hbox{\vrule width.#2pt height#1pt \kern#1pt
                   \vrule width.#2pt}
            \hrule height.#2pt}}}}

\def \cf{\frenchspacing cf.\ }

\def \cef{c_{\rm eff}}

%%%%%%%%%%%%%%%%%%%%%%%%%%%%%%%%%%%%%%%%%%%%%%%%%%%%%%%%%%%%%%%%%%%%%%
%%%%%%%%%%%%%%%%%%%%%%%%%%%%%%%%%%%%%%%%%%%%%%%%%%%%%%%%%%%%%%%%%%%%%%

{\nopagenumbers

\null
\line{\hfil preprint BONN-HE-92-36}
\line{\hfil hep-th/9211120}
\line{\hfil November 1992}
\vfill
\centerline {\grosss Lift of dilogarithm to partition identities}
\bigskip
\centerline{\phantom{mmmmmmmmmmmmmm}Michael Terhoeven}
\bigskip
\bigskip
\centerline{Physikalisches Institut}
\centerline{der}
\centerline{Rheinischen Friedrich-Wilhelms Universit\"at Bonn}
\centerline{Nussallee 12}
\centerline{53oo Bonn 1}
\centerline{unp044 at ibm.rhrz.uni-bonn.de}
\vfill
\centerline{\bf Abstract\phantom{mmmmmmmmmmmmmmmm}}
\medskip
For the whole set of dilogarithm identities found recently using the
thermodynamic Bethe-Ansatz for the $ADET$
series of purely elastic scattering theories we give partition
identities which involve characters of those conformal field theories
which correspond to the UV-limits of the scattering theories.
These partition identities in turn allow to derive the dilogarithm
identities using modular invariance and a saddle point approximation.
We conjecture on possible generalizations of this correspondance,
namely, a lift from dilogarithm to partition identities.
\bigskip
\vfill
\eject }

%%%%%%%%%%%%%%%%%%%%%%%%%%%%%%%%%%%%%%%%%%%%%%%%%%%%%%%%%%%%%%%%%%%%%
\pageno=1

\unter 1 Introduction

Recently there has been quite some interest in the dilogarithm function,
which satisfies a huge set of not obviously related identities.
In Nahm etal (1992) a method of Richmond and Szekeres (1981) to derive
dilogarithm identities from partition identities was resuscitated.
Furthermore, it was shown that the partitions in question can be
interpreted as characters of 2d conformal field theory (CFT) and that
the dilogarithm identities follow from equating the
asymptotic growth of the sum side of the partition identity to the
known growth of a character of a CFT given by the central charge.

The example considered in Nahm etal (1992) is the set of all $(2,odd)$
Virasoro minimal models, which corresponds to the tadpole series $T_n$
(sometimes called $A_{2n}^{(2)}$) of thermodynamic Bethe-Ansatz
(TBA) equations and dilogarithm identities in  Klassen, Melzer (1990).
Their $A_n$ series in turn is related to the $\zzz_n$ parafermion
conformal field theories and the corresponding partition identities
for the parafermion characters can be found in Lepowsky, Primc (1985)
(\cf the appendix for these two series).
The cases missing in the $ADET$ classification of Klassen, Melzer
(1990) of purely elastic scattering theories related to perturbations of
CFT are the exceptional ones $E_6, E_7$ and $E_8$, which correspond
to the $(6,7)$, $(4,5)$ and $(3,4)$ Virasoro minimal models, and the
$D_n$ series, which is believed to correspond to certain
$U(1)/\zzz_2$ orbifold
models. However, none of the partition identities in the mathematical
literature known to us (cf. Andrews (1976), (1986), Bressoud (1980))
do the
job, namely, equating the asymptotic growth of the partitions (which if
possible can be interpreted as characters of a CFT with appropriate
central charge) yields the dilogarithm identities of Klassen, Melzer
(1990).

One aim of this article is to close this gap: we will
write down in chapter 2 the
partition identities which give, using the method of Richmond, Szekeres
(1981), the $D_n^{(1)}$, $E_6$, $E_7$ and $E_8$ dilogarithm identities of
Klassen, Melzer (1990).
Indeed, as expected, the partitions can be interpreted as characters
of certain conformal field theories.

More than this result, we
would like to stress the way it was achieved: {\it Starting from a
dilogarithm identity {\rm with} corresponding TBA
equations, a sum side of a partition
identity was conjectured. Using computer algebra the other side was
identified with characters of a conformal
field theory}. In chapter 4 we present a conjecture on how this program
might generalize.

In chapter 3, we briefly comment on the invariance of the structure
of the partition identities under the tensor product of CFTs.

\medskip
\unter 2 The new identities

Let $C^{-1}_n$ be the inverse of the Cartan matrix $C_n$
of finite $D_n$ $(n\geq 3)$
$$
C^{-1}_n= \pmatrix{1& &\ldots &1 &1/2 &1/2    \cr
                      1&2&\ldots &2 &1 &1    \cr
        \vdots &\vdots &\ddots &\vdots &\vdots &\vdots  \cr
                      1&2&\ldots &n-2 &(n-2)/2&(n-2)/2  \cr
                    1/2&1&\ldots &(n-2)/2&n/4&(n-2)/4   \cr
                    1/2&1&\ldots &(n-2)/2&(n-2)/4&n/4   \cr }.
$$
Then the following identity seems to hold
$$
   q^{-1/24} \sum_{m_1,\ldots ,m_n\geq 0}
           { q^{{\bf m}C^{-1}_n {\bf m}^t} e^{\pi iz(m_n-m_{n-1})}
                 \over
              (q)_{m_1}\ldots (q)_{m_n} }
          \quad =\quad {1\over{\eta^{}(\tau)}}\
         \big( \Theta_{0,n} + \Theta_{n,n} \big)\ (z,\tau ,0),
\eqno(2.1)
$$
where $(q)_n = (1-q)\ldots(1-q^n)$ and the Jacobi theta functions
are given by
$$
 \Theta_{l,n} (z,\tau ,u) = e^{-2\pi iku}
         \sum_{m\in\zzz+l/2n} q^{nm^2} e^{-2\pi iznm}.
$$
A product representation of the partitions above can be found easily
from the theta function using the Jacobi tripel product identity
$$
\eqalign{
   \prod_{n=1}^\infty\
   (1-q^{2n}) (1+e^{2\pi iz} q^{2n-1}) (1+e^{-2\pi iz} q^{2n-1})
    & = \sum_{m=-\infty}^\infty q^{m^2}e^{2\pi imz},\cr
}
$$
namely
$$
\eqalign{
  & \Theta_{l,n}(z,\tau ,u) = \cr
        e^{-2\pi inu} q^{l^2/4n} e^{-\pi ilz}
        \prod_{m=1}^\infty \Big\lbrack &
        (1-q^{2mn})  (1+q^{(2m-1)n+l}e^{-2\pi inz})
        (1+q^{(2m-1)n-l}e^{2\pi inz})   \Big\rbrack . \cr
}
$$
We should remark that if $n$ is an integer multiple of $4$, the right
hand side of (2.1) can also be written as $\Theta_{0,n/4} (z,\tau)$.

Using the method of Richmond, Szekers (1981) and modular invariance
(cf. Nahm etal (1992))
one rederives from (2.1) at $z=0$ the following dilogarithm identity
given by Klassen, Melzer (1990)
$$
    {1\over L(1)} \bigg\lbrack
   2 L\bigg({1\over n-1}\bigg) +
   \sum_{a=2}^{n-1} L\bigg({1\over a^2}\bigg) \bigg\rbrack
     = 1 = c_{U(1)} .
\eqno(2.2)
$$
As expected, both, the dilogarithm identity (2.2)
as well as (2.1) show the connection to the $U(1)$ CFT with
central charge $c=1$.

\medskip
The corresponding partition
identities for the exceptional cases are given by
replacing the inverse of the Cartan matrix of $D_n$ on the left hand
side of (2.1) by the one of $E_6$, $E_7$ and $E_8$, setting $z=0$
and replacing the right hand side by
$\chi_{1,1} + \chi_{1,5} + 2\chi_{1,3}$, $\chi_{1,1} + \chi_{1,3}$
and $\chi_{1,1}$, where $\chi_{r,s}$ are the Virasoro
characters (\cf Rocha-Caridi (1984), Goddard etal (1986))
of the $(p,p')=(6,7)$, $(4,5)$ and $(3,4)$ unitary minimal models
with central charge $6/7$, $7/10$ and $1/2$ respectively
$$
   \chi^{(p,p')}_{r,s} (q) = {1 \over \eta^{}(q)}
   \big(       \Theta_{rp-sp',pp'} -
                \Theta_{rp+sp',pp'}    \big) (0,\tau ,0).
$$
For the corresponding dilogarithm identities we refer once again
to the paper of Klassen, Melzer (1990).

\medskip
\unter 3 Tensor products of CFT

It is remarkable that the structure of the identities of type (2.1)
is invariant under taking the product of two such identities:
One obtains an identity which involves the inverse Cartan matrices
$C^{-1}_1$, $C^{-1}_2$ of the single theories in the combination
$C^{-1}_1\oplus C^{-1}_2=$ $(C_1\oplus C_2)^{-1}$ on the left hand
side whereas on the right hand side we find characters of the tensor
product of the two CFTs (up to field identification).

This compatibility of the structure of the identities with
the tensor product of CFTs gives us additional
evidence for the conjecture in the following chapter.

\medskip
\unter 4 Lift of dilogarithm identities

Up to now we did not discuss another essential
structural element connecting
the partition and dilogarithm identities: the Bethe-Ansatz equations,
which appear in the method of Richmond, Szekeres (1981) as
saddle point conditions and
which have to be solved to obtain the arguments of the dilogarithm
functions. Also, these were conjectured in Nahm etal (1992) to
be closely related to the fusion algebras of the
corresponding CFTs.

In the cases considered in chapter 2 and the appendix, they take the
form
$$
       f_i = \prod_j (1-f_j)^{B_{ij}},
\eqno(4.1)
$$
where $B$ is two times the inverse of the Cartan matrix of any of the
$X_l$ diagrams with $X\in \{ A,D,E,T\}$ and $l$ the rank.
This can also be written in the suggestive form
$$
   B = C(A_1)\otimes C^{-1} (X_l)  ,
$$
where $C(A_1)=2$ is the Cartan matrix of $A_1=SU(2)$.

The suggestion made is that as a generalization
one should replace the Cartan matrix of $A_1$ by the Cartan matrix
of any Lie algebra. Indeed, as far as the Bethe-Ansatz equations
and corresponding dilogarithm identities are concerned,
this generalization
at least for $X=A$ and $A_1$ replaced by simply-laced algebras
does exist (Kirillov (1987), Kuniba, Nakanishi (1992),
Nahm etal (1992) and references therein).
We conjecture:

{\leftskip=1truecm
\noindent
Given any matrix $B=C(Y)\otimes C^{-1}(X)$, where $C(X)$, $C(Y)$
are the Cartan matrices corresponding to the Dynkin diagram of
$X,Y\in\{A,D,E,T\}$, the dilogarithm identities
corresponding to the Bethe-Ansatz equations (4.1) lift to
partition identities of the form
$$
   q^{-\cef /24} \sum_{m_1,\ldots ,m_n\geq 0}
           { q^{{\bf m}B {\bf m}^t/2 + {\bf b} {\bf m}^t}
                 \over
              (q)_{m_1}\ldots (q)_{m_n} }
          \quad =\quad \tchi_{\bf b}^{(X,Y)} (q) ,
$$
where $\bf b$ is a certain vector and $\tchi_{\bf b}^{(X,Y)}(q)$ a
linear combination with integer coefficients of the
characters of the CFT of effective central charge $\cef ={1\over L(1)}
\sum_i L(f_i)$, where the $f_i$ are the real solutions of (4.1).

}
\noindent
A few comments are in order:
In the case $X_l=A_l$ we would expect (cf. Kuniba, Nakanishi (1992))
that the CFT mentioned in the conjecture is the parafermionic theory
corresponding to the $Y_l$ WZW theory.

Concerning an extension of the conjecture to
non-simply-laced Lie algebras, we would expect that also
the denominators of the terms in the sum on the left hand side have
to be adjusted.

\medskip
\unter 5 Conclusion and Outlook

We believe that the mathematical structure encoded in the partition
identities conjectured above is the one given by Lepowsky
and Wilson (1981, 1984, 1985). For further possible applications in
mathematics physics we refer to the paper of Nahm etal (1992).

We hope that understanding better the conjecture will lead to a
better understanding of the general dilogarithm identities of
Kuniba, Nakanishi (1992) and to the creation of many new ones.

It is obvious that a first physical interpretation is bound to be
connected with perturbed CFT -- this being one recent occurrence of
dilogarithm identities. It seems that the structure of CFT knows
about its integrable perturbations in the sense that it allows
its characters to be written in a nice simple form which suggests
connections to integrable perturbations (compare the examples
$E_8$ and $A_1$ with Zamolodchikov (1989)).
However, the exact structural correspondance is unclear to us.
We refer to recent work of
Klassen, Melzer (1992), Kedem, McCoy (1992), Zamolodchikov (1991),
Ravanini, Tateo, Valleriani (1992) and references therein.

\medskip
\unter A \kern-6pt cknowledgements

I am indebted to W. Nahm, A. Recknagel, M. R\"osgen and R. Varnhagen
for essential discussions on related subjects.

At the time of writing I am supported by my parents and
the Studien\-stiftung des deutschen Volkes.

After this work was completed, it came to my attention
that a preprint (Kedem, Klassen, McCoy, Melzer (1992))
has recently appeared in which similar results were obtained.

\medskip
\unter A \kern-6pt ppendix

The case $T_n (=A_{2n}/\zzz_2)$ $(n\geq 1)$:

\noindent
The partition identities are special cases of the
Andrews-Gordon identities, which in turn
are certain generalizations of the well known Rogers-Ramanujan
identities. The one interesting for our application (\cf Andrews
(1976), Bressoud (1980)) is
$$
q^{-c^n_{\rm eff}/24}
    \sum_{m_1,\ldots ,m_{n}\geq 0}
           { q^{{\bf m}C^{-1}_n {\bf m}^t}
                 \over
             (q)_{m_1}\ldots (q)_{m_n} }
                           =  \chi^{(2,2n+3)}_{1,n} (q),
\eqno(A.1)
$$
where $C^{-1}_n$ is the inverse of the 'Cartan' matrix ($2\ -$ incidence
matrix) of the tadpole graph $T_n$ and $c^n_{\rm eff} = 2n/(2n+3)$
is the effective central charge of the $(2,2n+3)$ Virasoro minimal model.
A generalization of the above identity for all the characters of this
CFT exists. It involves in particular a
linear correction to the exponent of the numerator on the left hand side.

\noindent
The corresponding dilogarithm identity is given by
$$
   {1\over L(1)} \sum_{j=2}^{n+1}
        L\bigg({ \sin^2({\pi\over 2n+3}) \over
                \sin^2({\pi j\over 2n+3}) } \bigg)
        = c^n_{\rm eff} .
\eqno(A.2)
$$
The arguments of the dilogarithm can be identified with certain
quantum dimensions of the CFT.

\medskip
\noindent
The case $A_n$ $(n\geq 1)$:

\noindent
Here, the partition identities can be found in Lepowsky, Primc (1985)
$$
        q^{-c_n/24}
    \sum_{m_1,\ldots ,m_{n}\geq 0}
           { q^{{\bf m}C^{-1}_n {\bf m}^t}
                 \over
             (q)_{m_1}\ldots (q)_{m_n} }
     = (q)_\infty \sum_{m=0}^{n-1} c_m^0 (\tau ),
\eqno(A.3)
$$
where $C^{-1}_n$ is the inverse of the Cartan matrix of $A_n$
and $c^l_m$ are the so-called string functions at level $n+1$
(Kac, Peterson (1984))
$$
    c^l_m(\tau)\quad =\quad \eta(\tau)^{-3}\mkern-89mu \sum_{\matrix{
          \scriptstyle (x,y)\in\rrr^2\cr
          \scriptstyle-|x|<|y|\leq|x|\cr
          \scriptstyle (x,y)\ or\ (1/2-x,1/2+y)\in
           ((l+1)/2(k+2),m/2k)+\zzz^2 \cr}} \mkern-89mu
                     {\rm sign}(x)\ q^{(k+2)x^2-ky^2},
\eqno(A.4)
$$
(and $c^l_m(\tau)=0$ if $l\neq m \mod 2$) which
are the characters of the $\zzz_{n+1}$ parafermionic theory with
central charge $c_n=2n/(n+3)$. A similar representation exists
for all characters of the parafermionic theory.

\noindent
The corresponding dilogarithm identity is given by
$$
   {1\over L(1)} \sum_{j=2}^{n+1}
        L\bigg({ \sin^2({\pi\over n+2}) \over
                \sin^2({\pi j\over n+2}) } \bigg)
        = c_n .
\eqno(A.5)
$$

%%%%%%%%%%%%%%%%%%%%%%%%%%%%%%%%%%%%%%%%%%%%%%%%%%%%%%%%%%%%%%%%%%%%%
\medskip
\unter R \kern-6pt eferences

\baselineskip=12pt

\frenchspacing  % slightly larger spaces, see TeXbook p. 74

\bigskip
\parindent=-1truecm
\parskip=0.2cm
\leftskip=1truecm

G.E. Andrews (1976), {\sl The theory of partitions}, Encyclopedia
   of Mathematics and Its Applications, Vol. 2, Addison Wesley.

G.E. Andrews (1986), {\sl q-series: their development and application
   in analysis, number theory, combinatorics, physics, and computer
   algebra} (Conference Board of the Mathematical Sciences,
   Regional Conference Series in Mathematics, Number 66).

D.M Bressoud (1980), {\sl Analytic and combinatorical generalizations
   of the Rogers-Rama\-nu\-jan identities} (Memoirs of the American
   Mathematical Society, Number 227, Volume 24).

P. Goddard, A. Kent, D. Olive (1986), {\sl Comm. Math. Phys.} {\bf 103},
     105-119.

V. G. Kac, D. H. Peterson (1984), {\sl Advances in Math.} {\bf 53},
     125-264.

R. Kedem, T. R. Klassen, B. M. McCoy, E. Melzer (1992),
     {\sl Fermionic quasi-particle representations for characters
          of ${G_1\times G_1\over G_2}$}, preprint ITP-SB-92-64,
      RU-92-51, hep-th/9211102.

R. Kedem, B. M. McCoy (1992), {\sl Construction of modular branching
    functions from Bethe's equations in the 3-state Potts chain},
   Stony Brook preprint ITP-SB-92-56, hep-th.

A. N. Kirillov (1987), {\sl Zap. Nauch. Semin. LOMI} {\bf 164}, 121
     ((1989) {\sl J. Sov. Math.} {\bf 47}, 2450.

T. R. Klassen, E. Melzer (1990),
     {Purely elastic scattering theories and their ultraviolet limits},
     {\sl Nucl. Phys. }{\bf B338}, 485-528; (1992),
     {\sl Nucl. Phys. }{\bf B370}, 511.

A. Kuniba, T. Nakanishi (1992),
   {Spectra in conformal field theories from the Rogers dilogarithm},
    preprint hep-th/9206034, MRR-009-92, SMS-042-92.

Lepowsky, Primc (1985), {\sl Structure of the standard modules for
              the affine Lie algebra $A_1^{(1)}$}, AMS Series:
              Contemporary Mathematics, Vol. 46.

J. Lepowsky, R. Wilson (1981), {\sl Proc. Nat. Acad. Sci. U. S. A.}
      {\bf 78}, 7254-8; (1984), {\sl Invent. Math.} {\bf 77}, 199-290;
      (1985), {\sl ibid} {\bf 79}, 417-42.

W. Nahm, A. Recknagel, M. Terhoeven (1992), {\sl Dilogarithm
   identities in CFT}, hep-th/9211034, preprint BONN-HE-92-35.

F. Ravanini, R. Tateo, A. Valleriani (1992), {\sl Dynkin TBA's},
    Bologna preprint DFUB-92-11, Torino preprint DFTT-31/92,
    hep-th.

A. Rocha-Caridi (1984), {\sl Vertex operators in mathematics and
   physics}, ed. J. Lepowsky etal, MSRI publications No.3, Springer,
   p. 451.

M. R\"osgen, R. Varnhagen, J. Kellendonk (1992), in preparation.

B. Richmond, G. Szekeres (1981),
   {\sl J. Austral. Math. Soc.} {\sl (Series A) 31}, 362-373.

A. B. Zamolodchikov (1989), {\sl Adv. Stud. Pure Math.}, {\bf 19},
      641.

Al. B. Zamolodchikov (1991), {\sl Phys. Lett.} {\bf B253}, 391.

\bye